# Nonlinear Systems in Wireless Power Transfer Applications

H Chan, Tianjin University

*Abstract*—As a novel pattern of energization, the wireless power transfer (WPT) offers a brand-new way to the energy acquisition for electric-driven devices, thus alleviating the over-dependence on the battery. This report presents three types of WPT systems that use nonlinear control methods, in order to acquire an in-depth understanding of the course of Nonlinear Systems.

*Index Terms*—Wireless power transfer (WPT), nonlinear system, neural network.

## I. INTRODUCTION

NOWADAYS, nonlinear control methods are expanding rapidly. In particular, the neural network has become one of the most salient research areas in the past two decades. It has enormous advantages and has been successfully applied in many industrial fields. With its huge potential, power electronics benefit from the development of nonlinear control methods [1]. Regarding the wireless power transfer (WPT) technology, several nonlinear control strategies have been deployed in the corresponding WPT systems.

The rest of this report is organized as follows. Section II introduces a WPT system utilizing a conventional nonlinear control method, namely the idea of the negative resistance oscillator. Then, two advanced neural network-based control methods are discussed in Section III and Section IV, respectively. Finally, the conclusion of this report is put forward in Section V.

## II. NONLINEAR PARITY-TIME-BASED WPT SYSTEM

By introducing a concept of parity-time (PT) symmetry from quantum physics, a general nonlinear PT-symmetric model is proposed using a nonlinear negative resistance oscillator [2], which is shown in Fig. 1. It can attain stable power transfer with high and constant transfer efficiency under a dynamic change of the coupling condition. However, the series-series topology is more appropriate than the parallel-parallel topology due to its better capacity of power output. Accordingly, the series-series compensated WPT system is the focus of this section.

As shown in Fig. 2, the output characteristic of the inverter can be equivalent to a negative resistance by using a self-oscillating control strategy [3]. Then, the nonlinear PT-based WPT system can be realized without the wireless communication from the receiving side. Further, based on the coupled-mode theory, theoretical analyses show that the system achieves constant output power with constant efficiency in the strong coupling region against the continuous disturbance of mutual inductance, which can be well applied for in-flight wireless charging.

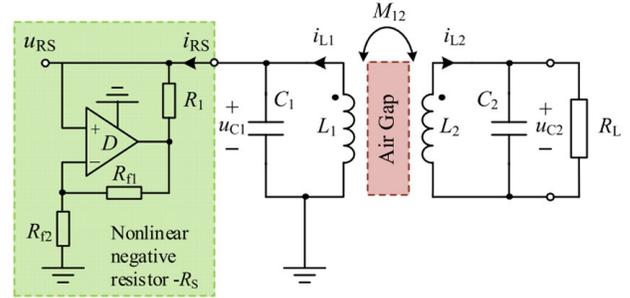

Fig. 1. Schematic of a nonlinear PT parallel-parallel WPT system [3].

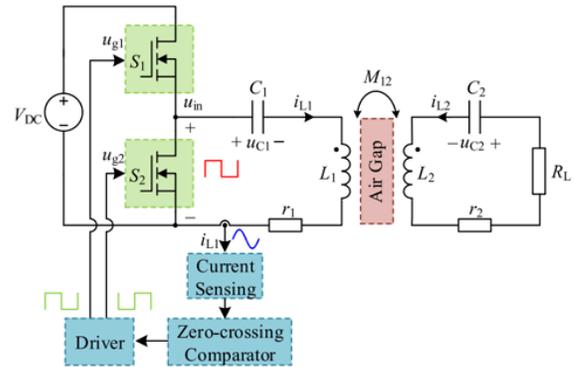

Fig. 2. Schematic of a nonlinear PT series-series WPT system [3].

An experimental prototype for drones is set up by adopting the proposed scheme, which can keep constant output power with a constant transfer efficiency of 93.6%. Fig. 3 depicts the experimental measured data of output power, transfer efficiency, and operating frequency. In the strong coupling region, the output power is constant near 10 W. The measured operating frequencies are also consistent with the calculated results based on the theoretical analysis in the article. Apparently, the nonlinear PT-based WPT system can practically provide stable power transfer with constant efficiency for the drone in a confined space, namely the strong coupling region.

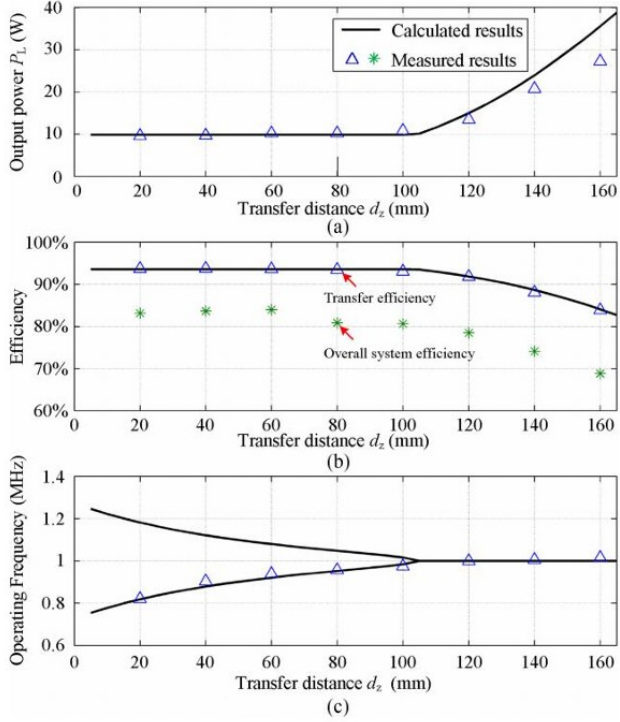

Fig. 3. Transfer performance of the prototype for the coaxial aligned case [3]. (a) The output power versus the transfer distance. (b) The transfer efficiency and overall system efficiency versus the transfer distance. (c) The operating frequency versus the transfer distance.

## III. BACKPROPAGATION NEURAL NETWORK-BASED CONTROL METHOD

Regarding the series-series compensated WPT system, a communication/model-free (C/M-free) constant current control scheme based on the feedforward-backpropagation neural network is proposed to deal with the continuous variation of the coupling effect while saving the communication link [4]. As the most significant difference from previous works, the proposed control scheme possesses salient advantages of the enhanced practicability by handling the continuous variation of the coupling effect, the real wireless by removing the communication link, and the reduced complexity by only employing primary current without any additional circuits.

Fig. 4 depicts the basic structure of the proposed C/M-free constant current control system. The primary current, which is acquired by the current sensor, is employed as one of the input signals for the backpropagation neural network. The training result is passed into the phase-shift controller that can produce phase-shift pulse width modulation (PWM) signals to adjust the output voltage of the H-bridge inverter. Thereby, the constant current output is realized with the control of the primary current through the proposed controller. Furthermore, the measurement and control are both implemented on the primary side, thus saving the bilateral communication successfully. Fig. 5 depicts the design approach and implementation process of the proposed C/M-free constant current control scheme.

In order to verify the feasibility of the proposed C/M-free constant current control, a self-regulating procedure is designed to imitate the continuous disturbance of the coupling effect in practical WPT systems. Fig. 6 depicts the variation of the output current in three different stages, including the initial stage (spacing between coils 12 cm), Stage I (spacing between coils 16 cm), and Stage II (spacing between coils 14 cm). The experimental results demonstrate that the proposed C/M-free constant current control scheme can effectively maintain the constant output current with a continuous fluctuation of the coupling effect between the transmitting and pickup coils.

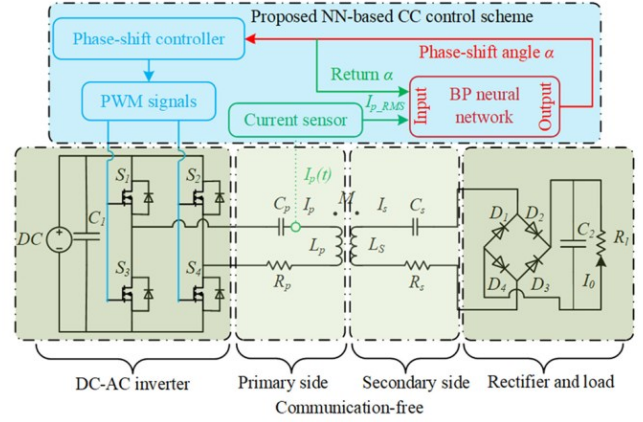

Fig. 4. Proposed C/M-free constant current control [4].

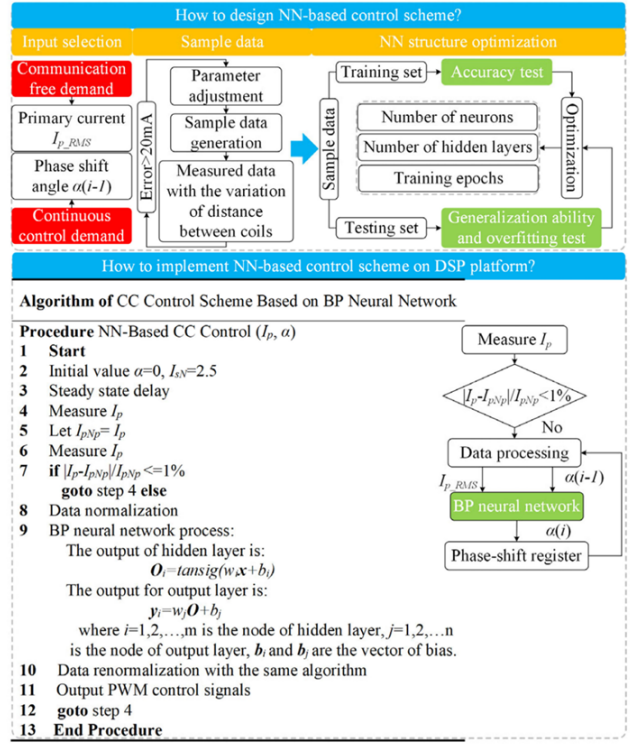

Fig. 5. Design and implementation of the proposed C/M-free constant current control [4].

Moreover, regarding the relatively high level of the load resistance fluctuation, this article proposes a robustness-enhanced solution to ensure the accuracy of the output current by taking the fluctuations of load resistance (5%) into account in the training data. The regulated results of the improved method are shown in Fig. 7(a). Fig. 7(b) shows the variation of the output current with respect to the fluctuation of the primary capacitance and inductance. When the fluctuations of both primary capacitance and inductance are within 1%, the decline of the output current is less than 0.1%, which has little effect on the control accuracy. However, as the fluctuations increase to 2-3%, the decrease of the output current is more obvious. Hence, the parameter dependence of the resonant capacitance and inductance is relatively high, which is a potential threat to be taken into account when utilizing the proposed C/M-free CC scheme. Therefore, the circuit components with low temperature drift are the preferable choice to decrease the fluctuations and further ensure control accuracy in practical applications. For instance, polypropylene film capacitors (KEMET R41) are chosen in this article, which can effectively avoid the impact of temperature drift on the accuracy of the output current.

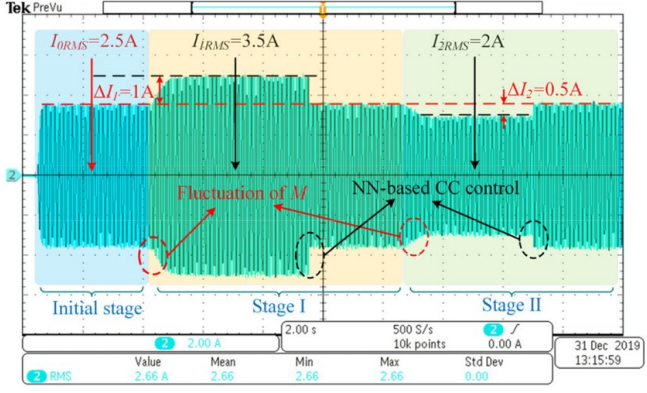

Fig. 6. Measured waveforms of the overall experimental test [4].

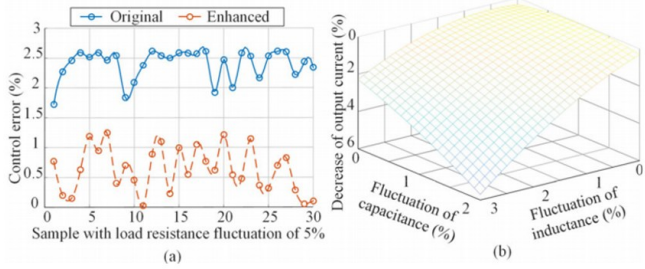

Fig. 7. Parameter sensitivity analysis [4]. (a) Control accuracy with respect to load fluctuation of 5%. (b) Decrease of the output current with respect to capacitance fluctuation within 2% and inductance fluctuation within 3%.

## IV. RADIAL BASIS FUNCTION NEURAL NETWORK-BASED CONTROL METHOD

Regarding the high-order (LCC-P) compensated WPT system, a dynamic-balancing robust constant current control is proposed for wireless in-flight charging systems by adopting the online-trained radial basis function neural network (RBFNN) [5]. The schematic of the proposed method is shown in Fig. 8, where the output current $I_L$ can be measured by the current sensor at the secondary side and then delivered to the digital signal processing (DSP) controller via a communication link.

In this report, the focus is on the implementation and the effectiveness of RBFNN. Considering the better generalization ability and the faster execution speed of the RBFNN, it is chosen as the structure of the neural network-based controller. Fig. 9 depicts the optimization of the proposed RBFNN-based control, where the online-trained RBFNN and the multidimensional optimization method are the key parts of the control algorithm. In order to meet the unique requirement of wireless in-flight charging systems, this article proposes the multidimensional optimization method as follows, including the center optimization with data augmentation, initial weight optimization, and error limit setting.

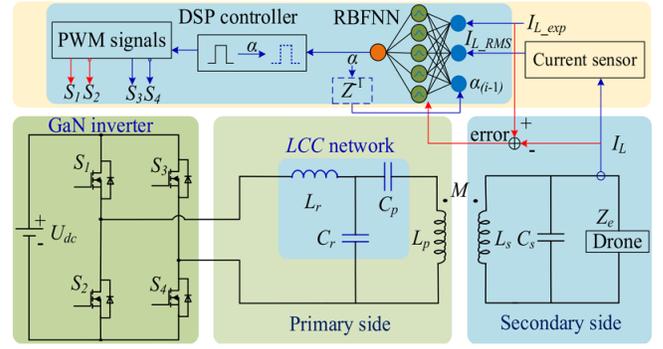

Fig. 8. Schematic of dynamic-balancing robust current control [5].

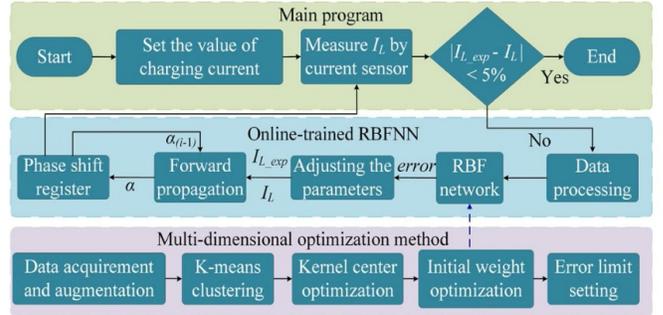

Fig. 9. Flowchart of dynamic-balancing robust current control [5].

1) Center optimization with data augmentation: First, the phase angle is arbitrarily adjusted to acquire the relationship of phase angle and output current under different coupling coefficients, and the data are recorded during the period of adjustment. Based on the recorded data, it is augmented according to (11) and (12). Then, the augmented data are used to select the partial center by the $K$-means clustering algorithm. In addition, because the expected current $I_{L\_\exp}$ is fixed, the center corresponding to the expected current is set as constant value. Accordingly, the center is preselected and partial adjusted during the parameter adjustment period, which saves the computation resources, thus increasing the execution speed of the proposed scheme.

2) Initial weight optimization: Because the RBFNN is online trained, the value of initial weight directly affects the training speed in the initial period. The optimized initial weight leads the RBFNN to learn the valuable information and prevents it from divergency. Hence, the initial value of weight is calculated as

$$\omega_i = ec_{\min} + i\frac{ex_{\max} - ec_{\min}}{2} \qquad (1)$$

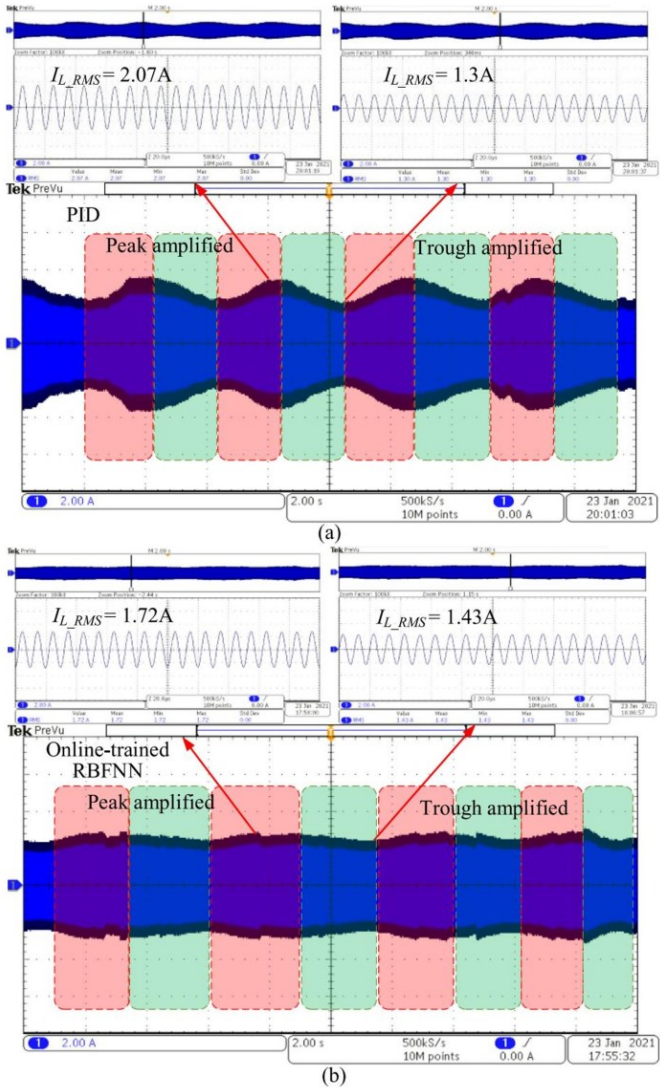

Fig. 10. Measured results of tracking continuously varied coupling effect in a period of 20 s [5]. (a) Conventional current control. (b) Proposed dynamic-balancing robust current control.

where $ec_{min}$ is the minimum value of the normalized phase angle in the generated data, $ec_{max}$ is the maximum value, and $i$ is the node number of the hidden layer.

3) Error limit setting: Even if under no disturbances, the output current may fluctuate within specified bounds. When the output current shifts from the expected value, the RBFNN-based control scheme is activated to restrain the error. However, the activation may cause the RBFNN to overfit in this data scope, namely, affect the generalization ability of RBFNN and, furthermore, affect the accuracy and rapidity of the network. Accordingly, the error limit setting is necessary to improve the stability and generalization of the network. This article set the error limit to 5%, which is a tolerance range for wireless in-flight charging systems.

After applying the multidimensional optimization method, the online-trained RBFNN is utilized as shown in the flowchart of the control scheme, where the parameters including the weights and partial center are adjusted by the backpropagation algorithm with momentum.

An experimental prototype is set up to validate the feasibility of the proposed dynamic-balancing robust current control. In particular, the comparative analysis is carried out with respect to the conventional PID control. For exemplification, a continuously varied coupling effect in a period of 20 s is designed, as shown in Fig. 10, and the expected output current is set to 1.6 A. In the exemplified period, the coupling effect varies between 0.102 and 0.157. Apparently, the online-trained RBFNN control significantly has a smaller tracking error than the PID control. The experimental result of such a 20-s period illustrates that the proposed dynamic-balancing robust current control possesses the ability to quickly track the continuously varied coupling effect by means of the property of partial response and generalization, which is the unique requirement of the wireless in-flight charging systems to deal with the continuous variation of coupling effect.

## V. CONCLUSION

With WPT as a background, this report outlines three high-level papers on nonlinear control methods. In particular, excellent control performance is shown against the unpredictable disturbance of the mutual inductance, both with the conventional method of the negative resistance oscillator and more advanced methods of neural networks. Apparently, nonlinear control methods are not only used in the WPT areas but also used in a wide range of industrial scenarios. As postgraduate students in the field of control, we need to pay more attention to the vital value of nonlinear control methods and advanced approximately linear models [6] in our future research. At the same time, regarding suitable applications, we should try to use nonlinear control methods as much as possible so as to enhance our understanding of them.

## ACKNOWLEDGMENT

The author would like to thank Prof. Wei for his great efforts in the course on Nonlinear Systems.

## REFERENCES

[1] K. Chen and Z. Zhang, "In-flight wireless charging: A promising application-oriented charging technique for drones," *IEEE Ind. Electron. Mag.*, vol. 18, no. 1, pp. 6–16, Mar. 2024.
[2] S. Assawaworrarit, X. Yu, and S. Fan, "Robust wireless power transfer using a nonlinear parity-time-symmetric circuit," *Nature*, vol. 546, no. 7658, pp. 387–390, Jun. 2017.
[3] J. Zhou, B. Zhang, W. Xiao, D. Qiu, and Y. Chen, "Nonlinear parity-time-symmetric model for constant efficiency wireless power transfer: Application to a drone-in-flight wireless charging platform," *IEEE Trans. Ind. Electron.*, vol. 66, no. 5, pp. 4097–4107, May 2019.
[4] Z. Zhang and W. Yu, "Communication/model-free constant current control for wireless power transfer under disturbances of coupling effect," *IEEE Trans. Ind. Electron.*, vol. 69, no. 5, pp. 4587–4595, May 2022.
[5] Z. Zhang, S. Shen, Z. Liang, S. H. K. Eder, and R. Kennel, "Dynamic-balancing robust current control for wireless drone-in-flight charging," *IEEE Trans. Power Electron.*, vol. 37, no. 3, pp. 3626–3635, Mar. 2022.
[6] K. Chen and Z. Zhang, "Rotating-coordinate-based mutual inductance estimation for drone in-flight wireless charging systems," *IEEE Trans. Power Electron.*, vol. 38, no. 9, pp. 11685–11693, Sept. 2023.